\documentclass[12pt]{article}
\usepackage{graphicx,color,latexsym,multirow,url,natbib}
\usepackage{ifthen,pifont,comment,graphicx,amssymb,amsthm,amsmath,enumerate}
\usepackage[dvipsnames]{xcolor}
\usepackage[letterpaper, left=3cm, top=3cm, right=3cm,
bottom=3cm]{geometry}
\RequirePackage[colorlinks,citecolor=blue]{hyperref}

 \makeatletter
\renewcommand\section{\@startsection {section}{1}{\z@}
                                   {-3.5ex \@plus -1ex \@minus -.2ex}
                                    {2.3ex \@plus.2ex}
                                   {\centering\normalsize\scshape}}
\renewcommand\subsection{\@startsection {subsection}{1}{\z@}
                                   {-3.5ex \@plus -1ex \@minus -.2ex}
                                   {2.3ex \@plus.2ex}
                                   {\centering\normalsize}}

\newtheorem{theorem}{Theorem}

\def\T{{ \mathrm{\scriptscriptstyle T} }}

\def\log{{\rm log}}

\def\tr{{\rm tr}}
\def\ud{{\rm d }}

\def\C{\small\textsc{C}}
\def\Ga{\small\textsc{Ga}}

\def\N{\small\textsc{N}}

\def\IB{\small\textsc{IB}}
\def\U{\small\textsc{U}}
\def\W{\small\textsc{W}}

\theoremstyle{plain}
\newtheorem{model}{{\itshape Model}}

\makeatother

\begin{document}
\begin{center}
{\large\bf On a Class of Shrinkage Priors for Covariance Matrix Estimation} \\
\vspace{.3in}
{ Hao Wang} \\
\medskip
{\it Department of Statistics, University of South Carolina, \\ Columbia, SC ~29208, U.S.A.} \\
 haowang@sc.edu
\bigskip

{Natesh S. Pillai} \\
\medskip
{\it Department of Statistics, Harvard University, \\ Cambridge, MA ~02138, U.S.A.} \\
pillai@fas.harvard.edu
\bigskip

This version: \today
\end{center}

\begin{abstract}
We propose a flexible class of models based on scale mixture of uniform distributions to
construct shrinkage priors for covariance matrix estimation. This new class
of priors enjoys a number of advantages over the traditional scale
mixture of normal priors, including its simplicity and flexibility in characterizing
the prior density. We also exhibit a simple, easy to implement Gibbs sampler for posterior
simulation
 which leads to efficient estimation
in high dimensional problems.  We first
discuss the theory and computational details of this new approach
and then extend the basic model
to a new class of multivariate conditional autoregressive models for
analyzing multivariate areal data. The proposed spatial model
flexibly characterizes both the spatial and the outcome correlation
structures at an appealing computational cost. Examples consisting of both
synthetic  and real-world data show the utility of this new
framework in terms of robust estimation as well as improved
predictive performance.
\end{abstract}

{\small \textit{Key words:} Areal data; Covariance matrix; Data augmentation Gibbs sampler; Multivariate conditional autoregressive model; Scale
mixture of uniform; Shrinkage; Sparsity.}

\section{Introduction}
Estimation of the covariance matrix $\Sigma$ of a multivariate
 random vector $y$
is ubiquitous in modern statistics and is
particularly challenging when the dimension of the covariance matrix,
$p$, is comparable or even larger than the sample size $n$.
For efficient inference, it is thus paramount to take advantage of
parsimonious structure often inherent in these high dimensional problems.  Many Bayesian approaches have been proposed for covariance matrix estimation by placing shrinkage priors on various parameterizations of the covariance matrix $\Sigma$. \citet{YangBerger1994} proposed reference priors for $\Sigma$ based on the spectral decomposition of $\Sigma$.   \citet{BarnardMcCullochMeng2000} and \citet{Liechty2004} considered shrinkage priors in terms of the correlation matrix and standard deviations. \citet{DanielsKass1999, DanielsKass2001} proposed flexible hierarchical priors based on a number of parameterizations of $\Sigma$. All of these methods use non-conjugate priors and typically rely on Markov chain algorithms which explore the state space locally such as Metropolis-Hastings methods or asymptotic approximations for posterior simulation and modeling fitting and are restricted to low-dimensional problems.

A large class of sparsity modeling of the covariance matrix
involves the identification of zeros of the inverse
$\Omega=\Sigma^{-1}$. This corresponds to the Gaussian graphical
models in which zeros in the inverse covariance matrix uniquely
determine an undirected graph that represents the strict conditional
independencies. The Gaussian graphical model approach for covariance
matrix estimation is attractive and has gained substantive
attention owing to the fact that its implied conditional dependence structure
provides a natural platform for modeling  dependence of random quantities
in  areas
such as biology, finance, environmental health and social sciences.
The standard Bayesian approach to inference in Gaussian graphical
models is the conjugate \textit{G}-Wishart prior
\citep{Roverato02,Massam05}, which places positive probability mass
at zero on zero elements of $\Omega$.  A zero constrained random
matrix $\Omega$ has the \textit{G}-Wishart distribution $\W_G(b,D)$
if its density is
\begin{eqnarray}
\label{eq:OmegaPrior} p(\Omega \mid G) &=& C_G(b,D)^{-1}
|\Omega|^{(b-2)/2} \exp \{-{1\over 2} \tr(D\Omega) \} \,1_{\{\Omega\in
M^+(G)\} },
\end{eqnarray}
where $b>2$ is the degree of freedom parameter, $D$ is a symmetric
positive definite matrix, $C_G(b,D)$ is the normalizing constant,
$M^+(G)$ is the cone of symmetric positive definite matrices with
entries corresponding to the missing edges of $G$ constrained to be
equal to zero, and $1_{\{\cdot\}}$ is the indicator function. Although
$G$-Wishart prior has been quite successfully used in many applications,
it has a few important limitations. First, the \textit{G}-Wishart prior is
sometimes not very flexible because of its restrictive form.  For example,
the parameters for the degrees of freedom are the same for all the elements of
$\Omega$. Second, unrestricted graphical model determination and
covariance matrix estimation is computationally challenging. Recent
advances for unrestricted graphical models
\citep{Jones05,WangCarvalho10,Mitsakakisetal,DLR2011} all rely on
the theoretical framework of \citet{Massam05} for sparse matrix
completion which is very computationally intensive.  Indeed, for non-decomposable
graphical models, we do not have a closed form expression
for the normalizing constant $C_G(b,D)$ and thus have to resort to tedious and often unstable Monte Carlo integration to estimate it for both graphical model determination and
covariance matrix estimation.

An alternative
method for Bayesian graphical model determination and estimation is
proposed by \citet{WongCarter03}. They placed point mass priors at
zero on zero elements of the partial correlation matrix and constant
priors for the non-zero elements. Their methodology applies to both
decomposable and non-decomposable models and is fitted by a
reversible jump Metropolis-Hastings algorithm. However, it is
unclear how to incorporate  prior information about individual entries
of $\Sigma$ in their
framework as the mathematical convenience of constant priors is essential for
their algorithm.

Absolutely continuous priors, or equivalently, penalty functions,
can also induce shrinkage to zero of subsets of elements of $\Omega$
and represent an important and flexible alternative to the point
mass priors. In the classical formulation, there is a rich
literature on methods for developing shrinkage estimators via
different penalty functions including the graphical lasso models
\citep{YuanLin2007,Friedman08,Rothman2008} and the graphical
adaptive lasso models \citep{FanFengWu2009} among many others. The
recent literature on Bayesian methods has focused on the posterior
mode estimation, with little attention on the key problem of
efficient inference on covariance matrix based on full posterior
computation, with the only exception of \citet{Wang2011} which gave
a fully Bayesian treatment of the graphical lasso models. One likely
reason is the difficulty in efficiently generating posterior samples
of covariance matrices under shrinkage priors.  A fully Bayesian
inference is quite desirable because it not only produces valid
standard errors and Bayes estimators based on decision-theoretic
framework but, perhaps more importantly, can be applied in multiple
classes of multivariate models that involve key components of
unknown covariance matrices such as the multivariate conditional
autoregressive models developed in Section \ref{sec:MCAR}.

This paper proposes a class of priors and the implied Bayesian
hierarchical modeling and computation for shrinkage estimation of
covariance matrices. A key but well known observation is that any symmetric, unimodal density
may be written as a scale mixture of uniform distributions.
 Our main strategy is to  use this mixture representations to construct
 shrinkage priors  compared to the traditional
methods for constructing shrinkage priors using the scale mixture of
normal distributions. As mentioned above, the scale mixture of uniform distribution is
not new to Bayesian inference. Early usage of this representation
includes Bayesian robust and sensitive analysis
\citep{Berger1985,BergerBerliner1986} and robust regressions with
heavy-tailed errors \citet{Walker97}.

However, our motivations are different;
we seek an approach for constructing tractable shrinkage
priors that are both flexible and computationally efficient.  We argue that the
class of scale mixture of uniform priors provide an appealing
framework for modeling a wide class of shrinkage estimation problems and
also has the potential to be  extended to a large class of
 high dimensional problems involving multivariate dependencies.
 We also highlight that a salient feature of our approach is its computational
simplicity. We construct a simple, easy to implement Gibbs sampler based on data augmentation for obtaining posterior draws
 for a large class of shrinkage priors. To the best of our knowledge, none of the existing Bayesian algorithms for sparse permutation invariant covariance estimation can be carried out solely based on a Gibbs sampler and they have to rely on Metropolis-Hastings methods. Since Gibbs samplers involve global proposal moves as compared to the local proposals of Metropolis-Hastings methods, in high dimensions this makes a difference in both the efficiency of the sampler and the running time of the algorithm.  Through simulation experiments, we
illustrate the robust performance of the scale mixture of uniform
priors for covariance matrix, as well as highlighting the strength
and weakness of this approach compared to those based on point
mass priors.   Through an extension to a class of multivariate
conditional autoregressive models, we further illustrate that the framework of
scale mixture of
uniforms naturally allows and encourages the integration of data and
expert knowledge in model fitting and assessment, and consequently
improves the prediction.

 The rest of the paper is organized as follows. In Section \ref{sec:2}
 we outline our framework for constructing shrinkage priors for covariance matrices
 using the scale mixture of uniforms.  In Section \ref{sec:smuPre} we construct a Gibbs sampler based on
  a data augmentation scheme for
 sampling from the posterior distribution. In Section \ref{sec:cov:simu} we conduct a simulation study and compare and contrast our models with existing methods.
  In Section \ref{sec:MCAR} we extend our model to build shrinkage priors on
 multivariate conditional autoregressive models. In Section \ref{sec:reg}  we briefly discuss the application of our methods for shrinkage estimation for regression models.

\section{Shrinkage priors for precision matrices} \label{sec:2}
\subsection{Precision matrix modeling}
Let $y
=(y^{(1)},y^{(2)},\ldots,y^{(p)})^\T$ be a $p$-dimensional random
vector having a multivariate normal distribution $\N(0,\Sigma)$ with
mean zero and covariance matrix $\Sigma$.  Let $\Omega =
(\omega_{ij})_{p\times p} =\Sigma^{-1}$ denote the precision matrix, i.e.,
 the inverse of the
covariance matrix $\Sigma$. Given a set of independent random samples
$Y=(y_1,\ldots,y_n)_{p\times n}$ of $y$, we wish to estimate the matrix $\Omega$.

We consider the following prior distribution for the precision matrix:
\begin{eqnarray} \label{cov:prior1} p(\Omega \mid \tau) &\propto& \prod_{i\leq j }
g_{ij}({\omega_{ij}-m_{ij}\over \tau_{ij}})\,1_{\Omega \in M^+},
\end{eqnarray}
 where  $g_{ij}(\cdot)$ is a
continuous, unimodal and symmetric probability density function with mode zero on $\mathbb{R}$, $M^+$ is the
space of real valued symmetric, positive definite $p \times p$ matrices, $\tau_{ij}>0$ is a scale
parameter controlling the strength of the shrinkage and $1_{A}$ denotes the indicator function of the set $A$. Our primary motivation for
constructing prior distributions of the form \eqref{cov:prior1} is that, often in real applications
the amount of prior information the modeler can vary across individual elements of $\Omega$.  For instance, one might incorporate the information that the variance of certain entries of $\Omega$ are close to $0$, or constrain some entries to be exactly $0$.
In this setting, shrinking different elements of $\Omega$ at a different rate clearly provides a flexible framework for conducting Bayesian inference. In addition to obtaining a flexible class of prior distributions,  by using a mixture representation for the density $g_{ij}$, we can construct a simple, efficient and easy to implement Gibbs sampler
to draw from the posterior distribution of the precision matrix.

\subsection{Scale mixture of uniform distributions}\label{sec:prior:example}
Our main tool is the following theorem which says that
all unimodal, symmetric densities may be expressed as scale mixture
of uniform distributions. 
\begin{theorem}\citet{Walker97,Feller-1971} \label{therm1}
Suppose that $\theta$ is a real-valued random quantity with a
continuous, unimodal and symmetric distribution with mode zero having density
$\pi(\theta) \, (-\infty<\theta<\infty)$. Suppose
$\pi^\prime(\theta)$ exists for all $\theta$. Then $\pi(\theta)$ has the
form:
\begin{eqnarray}\label{eq:unifmixture}
\pi(\theta) & = & \int_{0}^\infty {1\over 2 t}
\,1_{\{|\theta|<t\}}\,h(t)\,\ud t,
\end{eqnarray}
where $h(t) \propto -2t \times \pi^\prime(t)$ is some density function
on $[0,\infty)$. Therefore we may write
$$
\pi(\theta \mid t) \sim \U(-t,t),\quad
h(t) \propto - 2t \times \pi^\prime(t). 
$$
\end{theorem}
The generality and simplicity of Theorem \ref{therm1} allow us to
characterize various shrinkage priors by using the special structure
of (\ref{eq:unifmixture}). Indeed, as noted in \citet{Walker97}, a Gaussian
random variable $x \sim \N(\mu,\sigma^2)$ can be expressed as
$x \mid v \sim \U(\mu-\sigma\surd v, \mu + \sigma \surd v), v \sim \Ga(3/2, 1/2),$
which shows that, all of the distributions which
may be written as a scale mixture of Gaussian distributions can indeed
be expressed as a scale mixture of uniform distributions as well.
Let us discuss a few more examples of popular shrinkage priors
where Theorem \ref{therm1} is applicable.

A popular class of distributions for constructing shrinkage priors is the exponential power family given by $\pi(\theta) \propto \exp(-|\theta|^q/\tau^q),$
where the exponent $q>0$ controls the decay at the tails.
The mixing density function $h(t)$ given in \eqref{eq:unifmixture} can be thought
of as the ``scale" parameter.
In this case  we have $h(t)\propto t^q \exp(-t^q/\tau^q),$
which corresponds to the
generalized gamma distribution. Two important special cases are the
Gaussian distribution ($q=2$), and the double-exponential distribution ($q=1$), which have been
studied extensively in the context of the Bayesian lasso regression
\citep{ParkCasellla2008,Hans2009} and the Bayesian graphical lasso
\citep{Wang2011}. For general $q>0$, one may write the
exponential power distribution as a scale mixture of Gaussian distributions
\citep{Andrews1974,West1987}.  However, a
fully Bayesian, computationally efficient
 analysis is not  available based on  Gaussian
 mixtures, especially in the context of covariance estimation and
 graphical models. A few approximate methods exist
 for doing inference using the exponential power prior distribution such as the
variational method proposed by \citet{Armagan09}. Our use
of uniform mixture representation has the advantage of posterior simulation
via an efficient Gibbs sampler for any $q>0$ as is shown in Section
\ref{sec:samplealg} and further exemplified in Sections
\ref{sec:cov:simu} and \ref{sec:reg}.

Another natural candidate for shrinkage priors is the Student-\textit{t} distribution given by
$\pi(\theta)\propto(1+\theta^2/
\tau^2)^{-(\nu+1)/2},$ for which it is easy to show that $h(t) \propto t^{2} (1+
t^2/\tau^2 )^{-(\nu+3)/2}$. Hence, $t^2/\tau^2$ is an inverted beta distribution $\IB(3/2,\nu/2)$.
Recall that the inverted beta distribution $\IB(a,b)$ has the density given by $ p(x) \propto x^{a-1} (1+x)^{-a-b}1_{x>0}$.

The generalized double Pareto distribution is given by
$ \pi(\theta) \propto (1+|\theta|/
\tau)^{-(1+\alpha)},$
which corresponds to $h(t) \propto
t(1+t/\tau)^{-(2+\alpha)}$; i.e., the scale $t/\tau$ follows an inverted beta
distribution $\IB(2,\alpha)$. \citet{Armagan11} investigated the
properties of this class of shrinkage priors.

The above discussed class of shrinkage priors are well known
and documented. In the following we give a new distribution which we call
the ``logarithmic'' shrinkage prior
which seems to be new in the context of shrinkage priors. Consider
the density given by
\begin{eqnarray}\label{eq:Log}\pi(\theta) &\propto&
\log(1+\tau^2/\theta^2)\;.\end{eqnarray}
It is easy to show that the corresponding mixing distribution has
the half-Cauchy density,
$$h(t) \propto
(1+t^2/\tau^2)^{-1} 1_{\{t>0\}} \;.$$
  This prior has two desirable
properties for shrinkage estimation:
an infinite spike at zero and heavy tails.
These are precisely the desirable characteristics
of a shrinkage prior distribution as argued convincingly for
the ``horseshoe'' prior in \citet{CarvalhoPolsonScott2010}. The
horseshoe prior is constructed by scale mixture of normals, namely,
$\theta\sim \N(0,\sigma^2), \sigma \sim \C^+(0,1),$ where
$\C^+(0,1)$ is a standard half-Cauchy distribution on the
positive reals with scale 1. The horseshoe prior does not have a
closed form density but satisfies the following:
$$ {K\over 2} \log(1+4/\theta^2)<\pi(\theta)< K \,\log(1+2/\theta^2),$$
for a constant $K > 0$.
Clearly, our new prior (\ref{eq:Log}) has identical behavior
at the original and the tails as that of the horseshoe prior distribution
with the added advantage of having an explicit density function
unlike the horseshoe prior.

\subsection{Posterior sampling}\label{sec:samplealg}
Let $y$ denote the observed data. The scale mixture of uniform
representation provides a simple way of sampling from the posterior
distribution $p(\theta \mid y) \propto f(y\mid \theta)\pi(\theta)$,
where $f(y\mid \theta)$ is the likelihood function and $\pi(\theta)$ is the
shrinkage prior density. The representation (\ref{eq:unifmixture}) leads
to the following full
conditional distributions of $\theta$ and $t$ (conditional on $y$) given by
\begin{equation}\label{eqn:fullcondgibbs}
p(\theta \mid y,t)\propto f(y \mid \theta)\,1_{|\theta|<t},\quad p(t \mid y,\theta)\propto -\pi^\prime(t)\,1_{
|\theta|<t} \;.
\end{equation}
Thus the data augmented Gibbs sampler for obtaining posterior draws from
$p(\theta,t\mid y)$ involves iteratively simulating
from the above two conditional distributions. Simulation of the former involves
sampling from a truncated distribution, which is often achieved by
breaking it down further into several Gibbs steps, while sampling
the latter is achieved by the following theorem.

\begin{theorem}\label{therm:sampler}
Suppose the shrinkage prior density $\pi(\theta)$ can be represented by a
scale mixture of uniform as in equation (\ref{eq:unifmixture}). Then
the (posterior) conditional probability density function of the latent
scale parameter $t$ is given by
$$p(t \mid y,\theta)\propto -\pi^\prime(t)\, 1_{t
>|\theta|},
$$
and the corresponding (conditional) cumulative distribution function is
\begin{eqnarray}\label{cdf:t} u=F(t \mid y,\theta) = \textrm{pr}(T < t \mid y,\theta) &=& {\pi(|\theta|) -
\pi(t)\over \pi(|\theta|)} \quad |\theta|<t\;.\end{eqnarray}
\end{theorem}
The advantage of the above theorem is that it gives an explicit
expression of the conditional cumulative distribution function  in
terms of the prior density $\pi(\cdot)$. This provides a simple way to
sample from $p(t\mid y,\theta)$ using the inverse cumulative
distribution function method whenever $\pi(\cdot)$ can be easily
inverted. Table \ref{tab:priorexample} summarizes the density
functions of $\pi(\theta)$ and $h(t)$, and the inverse conditional
cumulative distribution function $F^{-1}(u\mid y,\theta)$ for
several shrinkage priors introduced in Section
\ref{sec:prior:example}. We note that the scale mixture of uniform
distributions are already used for doing inference for regression models using
the Gibbs sampler outlined above, for instance see \citet{Walker98}.

\begin{table}[tbp]
\caption{Density of $\theta$ and $t$ for some common
shrinkage prior distributions, along with the conditional posterior inverse
cumulative probability function for sampling $t$. Densities are
given up to normalizing constants. }
\footnotesize
\begin{tabular}{llll}
Density name & Density for $\theta$ & Density for $t$ & Inverse CDF:
$F^{-1}(u \mid \theta)$ \\
Exponential power  & $\exp(-|\theta|^q/\tau^q)$ & $t^q \exp(-
t^q/\tau^q)$ & $\{-\tau^q(\log\, u) +|\theta|^q\}^{1/q}$\\
Student-\textit{t} & $(1+\theta^2/\tau^2)^{-(\nu+1)/2}$&$t^{2} (\nu+t^2/\tau^2)^{-(\nu+3)/2}$&
$\{u^{-2/(\nu+1)}(\tau^2+\theta^2)-\tau^2 \}^{1/2}$
\\Generalized double Pareto & $(1+|\theta|/\tau)^{-(1+\alpha)}$ &
$t(1+t/\tau)^{-(2+\alpha)}$& $u^{-1/(1+\alpha)}(|\theta|+\tau)-\tau$\\
Logarithmic& $\log(1+\tau^2/\theta^2)$ & $ (1+t^2/\tau^2)^{-1}$ &
$\tau \{(1+\tau^2/\theta^2)^{u}-1\}^{-1/2}$
\end{tabular}
\label{tab:priorexample}
\\
CDF, cumulative distribution function.
\end{table}

\section{Posterior computation for precision matrices}\label{sec:smuPre}
 \subsection{Gibbs sampling on given global shrinkage parameter
$\tau$}\label{sec:fixtau}
Recall that given a set of independent random samples
$Y=(y_1,\ldots,y_n)_{p\times n}$ from a multivariate normal distribution $\N(0,\Omega^{-1})$, we wish to estimate the matrix $\Omega$ using the prior distribution given by (\ref{cov:prior1}). Let $T = \{t_{ij}\}_{i \geq j}$ be the
vector of latent scale parameters. For simplicity we  first consider a simple case where
$g_{ij}(\cdot)=g(\cdot), m_{ij}=0$ and $\tau_{ij}=\tau$ in this section, and then discuss the strategies for choosing
$\tau$ in Section \ref{sec:global}. However our algorithms can be
easily extended to the general case of unequal shrinkage parameters $\tau_{ij}$. 
Theorem \ref{therm1} suggests that the prior (\ref{cov:prior1})
can be represented as follows: $$p(\Omega \mid \tau) =
\int_{T}p(\Omega,T \mid \tau) \ud T \propto \int_{T}\prod_{i \geq j}
\big[ {1\over 2t_{ij}}1_{\{|\omega_{ij}|<\tau t_{ij}\}}
h(t_{ij})\big]\ud T,$$ where $p(\Omega,T \mid \tau)\propto \prod_{i
\geq j} \big[ {1/(2t_{ij})}\,1_{\{|\omega_{ij}|<\tau t_{ij}\}}
h(t_{ij})\big]$ is the joint prior and $h(t_{ij})\propto -t_{ij}
g^\prime(t_{ij})$. The joint posterior distribution of $(\Omega,T)$
is then:
\begin{eqnarray} p(\Omega, T \mid Y,\tau) &\propto& |\Omega|^{n/2} \exp \{-{1\over 2}\tr(S \Omega) \}
\prod_{i \geq j}\big[ -1_{\{|\omega_{ij}|<\tau t_{ij}\}}\,
g^\prime(t_{ij})\big] \label{cov:post1},
\end{eqnarray}
where $S = YY^\T$.

 The most direct approach for sampling from
(\ref{cov:post1}) is to update each $\omega_{ij}$ one at a time
given the data, $T$, and all of the entries in $\Omega$ except for
$\omega_{ij}$ in a way similar to those proposed in
\citet{WongCarter03}. However, this direct approach requires a separate
Cholesky factorization for updating each $\omega_{ij}$ to find its
allowable range and conditional distribution. It also relies on
the Metropolis-Hastings step to correct the sample. We describe an
efficient Gibbs sampler for sampling $(\Omega,T)$ from
(\ref{cov:post1}) that involves one step for sampling $\Omega$ and
the other step for sampling $T$.

Given $T$, the first step of our Gibbs sampler systematically scans
the set of $2\times 2$ sub-matrices $\{\Omega_{e,e}: e=(i,j), 1 \leq
j<i \leq p \}$ to generate $\Omega$.  For any $e=(i,j)$, let
$V=\{1,\ldots,p\}$ be the set of vertices and note that
$$|\Omega|=|A||\Omega_{V\backslash e,V\backslash e}|,$$ where $A$, the
Schur component of $\Omega_{V\backslash e,V\backslash e}$, is
defined by $A= \Omega_{e,e}- B$ with $B=\Omega_{e,V\backslash e}
(\Omega_{V\backslash e,V\backslash e})^{-1} \Omega_{V\backslash
e,e}$. The full conditional density of $\Omega_{e,e}$ from
(\ref{cov:post1}) is given by
$$p(\Omega_{e,e} \mid -)\propto |A|^{n/2}\exp\{-{1\over 2} S_{e,e} A \}\,1_{\{\Omega_{e,e}\in \mathcal{T}\}},$$
where
$\mathcal{T} = \{|\omega_{ij}|<\tau t_{ij}\}\cap\{|\omega_{ii}|<\tau t_{ii}\}
\cap \{|\omega_{jj}|<\tau t_{jj}\}$. Thus,  $A$ is a truncated Wishart
variate. To sample $A$, we write $$A = \left(\begin{array}{cc}
                               1 & 0 \\
                               l_{21} & 1 \\
                             \end{array}
                           \right) \left(\begin{array}{cc}
                               d_1 & 0 \\
                               0 & d_2 \\
                             \end{array}
                           \right)\left(\begin{array}{cc}
                               1 & l_{21} \\
                               0 & 1 \\
                             \end{array}
                           \right),\quad S_{e,e}=\left(\begin{array}{cc}
                               s_{11} & s_{12} \\
                               s_{21} & s_{22} \\
                             \end{array}
                           \right),$$
with $d_1>0$ and $d_2>0$.  The joint distribution of
$(l_{12},d_1,d_2)$ is then given by: $$ p( d_1,d_2,l_{21} \mid -) \propto
d_1^{n/2+1}d_2^{n/2} \exp[-{1\over 2}\tr
\{s_{11}d_1+s_{22}(l_{21}^2d_1+d_2)+2s_{21}d_1l_{21}\}
]\,1_{\Omega_{e,e}\in \mathcal{T}}\label{cov:post2},
$$
which implies that the univariate conditional distribution for the parameters
$d_1$ and $d_2$ is a truncated
gamma distribution, and a truncated normal distribution for $l_{21}$.
Details of the parameters of the truncated region and strategies for
sampling are given in the Appendix.  Given $\Omega$, the second step
of our Gibbs sampler generates $T$ in block using the inverse cumulative distribution function
methods described in equation (\ref{cdf:t}). These two steps
complete a Gibbs sampler for model fitting under a broad class of
shrinkage priors for $\Omega$.


One attractive feature of the above sampler is that it is also
suitable for sampling $\Omega \in M^+(G)$, that is, $\Omega$ is
constrained by an undirected graph $G=(V,E)$ where $V$ is the set of
vertices and $E$ is  a set of edges and $\omega_{ij}=0$ if and only
if $(i,j)\notin E$. The ability to sample $\Omega \in M^+(G)$ is
useful when substantive prior information indicates a certain subset
of elements in $\Omega$ are indeed zero.  Section \ref{sec:MCAR}
provides such an example that involves a class of multivariate spatial
models. To sample $\Omega \in M^+(G)$,  the only modification
that is required is to replace the set of all $2\times 2$ sub-matrices
$\{\Omega_{e,e}: e=(i,j), 1 \leq j< i \leq p \}$  with the set
$\{\Omega_{e,e}: e \in E \}\cup \{\Omega_{v}: v \in V_I\}$ where
$V_I$ is the set of isolated nodes in $G$.


\subsection{Choosing the shrinkage parameters}\label{sec:global} We start with the scenario when $\tau_{ij}=\tau$ and $m_{ij} = 0$ for all $i\geq j$. In this case we have
$$p(\Omega \mid \tau) = C_\tau^{-1} \prod_{i \geq j}
g({\omega_{ij}\over \tau}),$$ where $C_\tau$ is a normalizing term
involving $\tau$. This normalizing constant
is a necessary quantity for choosing hyper parameters for
$\tau$.  Since $p(\Omega \mid \tau)$ is a scale family, applying the substitution
$\tilde{\Omega} = \Omega/\tau$ yields,
\begin{equation} \label{eq:constanttau} C_\tau = \int_{\Omega\in M^+}\prod_{i \geq j}
g({\omega_{ij}\over \tau})\ud \Omega  = \tau^{{p(p+1)\over
2}}\int_{\tilde{\Omega}\in M^+} g(\tilde{\omega}_{ij})\ud
\tilde{\Omega},
\end{equation}
where the integral on the right hand side of the above equation  does not involve $\tau$ because $\{\tilde{\Omega}:\tilde{\Omega}
\in M^+\} = \{\Omega:\Omega \in M^+\}$. Hence, under a hyperprior
$p(\tau)$, the conditional posterior distribution of $\tau$ is
\begin{equation}\label{eqn:condposttau}
p(\tau \mid Y,\Omega) \propto \tau^{-p(p+1)/2}\prod_{i \geq j}
g({\omega_{ij}\over \tau}) p(\tau)\;.
\end{equation}
 Now the sampling scheme in
Section \ref{sec:fixtau} can be extended to include a component to
sample $\tau$ at each iteration.

Now suppose $m_{ij} = 0$ and instead of having a single global shrinkage parameter, we wish to control the rate at which the individuals elements of $\Omega$ are shrunk towards $0$ separately. A natural shrinkage prior for this problem is
$$p(\Omega \mid \tau) = C_\tau^{-1} \prod_{i \geq j}
g_{ij} ({\omega_{ij}\over \tau})$$
where $g_{ij}$ may all be different. The idea is that by choosing a different density $g_{ij}$ for
each edge, we can incorporate the prior information for the rate at which different entries of $\Omega$ are shrunk towards $0$. For a hyper prior $p(\tau)$, using an identical calculation as in \eqref{eq:constanttau} and \eqref{eqn:condposttau} we deduce that the conditional posterior of $\tau$ is then given by
\begin{equation}
p(\tau \mid Y,\Omega) \propto \tau^{-p(p+1)/2}\prod_{i \geq j}
g_{ij}({\omega_{ij}\over \tau}) p(\tau)\;.
\end{equation}
Notice that the Gibbs sampler presented in  Section \ref{sec:fixtau} applies to this case as well; we just need to use the cumulative distribution function for the density $g_{ij}$ for sampling from the conditional distribution of $t_{ij}$. Alternatively, one can also fix a density $g$ and write
$p(\Omega \mid \tau) = C_\tau^{-1} \prod_{i \geq j}
g ({\omega_{ij}\over v_{ij}\tau})$ for fixed positive constants $v_{ij}$ and then make inference about the common $\tau$.

We conclude this section with the remark  that our approach can be adapted for hierarchical models. For example, in Section \ref{sec:MCAR} we
consider a shrinkage prior that shrinks $\Omega$ towards a
given matrix $M=(m_{ij})$ under the constraint that
$\Omega\in M^+(G)$ for a given graph $G$:
$$ p(\Omega) = C_{\tau,M}^{-1} \prod_{(i,j)\in E}
g({\omega_{ij}-m_{ij} \over \tau})1_{\Omega \in M^+(G)},$$ where $E$ denotes
the set of edges of the graph $G$ and
normalizing constant $C_{\tau,M} = \int_{\Omega\in
M^+(G)}\prod_{(i,j)\in E} g({\omega_{ij}-m_{ij}\over \tau})\ud
\Omega$ is the normalizing constant. In this case $C_{\tau,M}$ is analytically intractable as a function of $\tau.$  In the
example of Section \ref{sec:MCAR},  we fixed $\tau$ at a value that
represents prior knowledge of the distribution of $\Omega$ to avoid
modeling $\tau$.  In some applications, it may be desirable to add another
level of hierarchy for modeling $\tau$ so that we can estimate it from data. Several
approaches have been proposed for dealing with the intractable normalizing constant, see
\citet{Liechty2004}, \citet{Liechty2009} and the references therein for one such approach.

\section{Simulation experiments} \label{sec:cov:simu}
To assess the utility of the scale mixture of uniform priors, we
compared a range of priors in this family against three
alternatives: the frequentist graphical lasso method of
\citet{Friedman08}, the Bayesian \textit{G}-Wishart prior and the
method of \citet{WongCarter03}. The latter two place positive prior
mass on zeros. We considered four covariance matrices from
\citet{Rothman2008}:

\begin{model}
An \textsc{AR}(1) model with $\sigma_{ij}=$0$\cdot$7$^{|i-j|}$.
\end{model}

\begin{model}
 An \textsc{AR}(4) model with
$\omega_{ii}=$1, $\omega_{i,i-1}=\omega_{i-1,i}=$0$\cdot$2, $\omega_{i,i-2}=\omega_{i-2,i}=\omega_{i,i-3}=\omega_{i-3,i}=$0$\cdot$2, $\omega_{i,i-4}=\omega_{i-4,i}=$0$\cdot$1.
\end{model}

\begin{model}
A sparse model with $\Omega = B + \delta
I_p$ where each off-diagonal entry in $B$ is generated independently
and assigned the value 0$\cdot$5 with probability $\alpha=$0$\cdot$1 and 0 otherwise.  The diagonal elements $B_{ii}$ are set to be 0, and $\delta$ is chosen so that
the condition number of $\Omega$ is $p$. Here the condition number
is defined as $\max(\lambda)/\min(\lambda)$ where $\max(\lambda), \min(\lambda)$ respectively denote the maximum and minimum eigenvalues of the matrix $\Omega$.
\end{model}
\begin{model}
A dense model with the same $\Omega$ as in model 3 except for
$\alpha=$0$\cdot$5.
\end{model}
For each of the above four models, we generated samples of size $n = 30,100$  and
dimension $p = 30$, yielding  the
proportion of non-zero elements to be  $6\%, 25\%,10\%,50\%$, respectively.
We compute the risk under two standard loss functions, Stein's loss function,
$L_1(\hat{\Sigma},\Sigma)=\tr(\hat{\Sigma}\Sigma^{-1})-\log(\hat{\Sigma}\Sigma^{-1})-p$,
and the squared-error loss function $L_2(\hat{\Sigma},\Sigma)
=\tr(\hat{\Sigma}-\Sigma)^2$. The corresponding Bayes estimators are
$\hat{\Sigma}_{L_1}=\{\rm{E}(\Omega\mid Y)\}^{-1}$ and
$\hat{\Sigma}_{L_2}= \rm{E}(\Sigma \mid Y)$, respectively. We used
the posterior sample mean using the Gibbs sampler for estimating
the risk for the Bayesian methods and the
maximum likelihood estimate for the graphical lasso
method.

When fitting graphical lasso models, we used the 10-fold
cross-validation to choose the shrinkage parameter. When fitting the
\textit{G}-Wishart priors, we followed the conventional prior
specification $\Omega \sim \small{\textsc{W}}_G(3,I_p)$ and used the
reversible jump algorithm of \citet{DLR2011} for model fitting. For
both the \textit{G}-Wishart priors and the methods of
\citet{WongCarter03}, we used the default graphical model space
prior \citep{CarvalhoScott2009} $$p(G) = \{(1+m){m \choose |G|}\}^{-1},$$ where $m=p(p-1)/2$ and
$|G|$ is the total number of edges in graph $G$. For the scale
mixtures of uniforms, we considered the exponential power prior $p(\Omega\mid \tau)\propto \exp\{-\sum_{i\leq j} |\omega_{ij}|^q/\tau^q\}$
with $q\in \{$0$\cdot$2, 1$\}$, the generalized
double-Pareto prior $p(\Omega\mid \tau) \propto \prod_{i\leq j}(1+|\omega_{ij}|/\tau)^{-1-\alpha}$ and the new logarithmic
prior $p(\Omega\mid \tau) \propto \prod_{i\leq j}\log(1+\tau^2/\omega_{ij}^2)$. For the choice of the global shrinkage
parameters, we assumed the conjugate distribution $\tau^{-q} \sim
\small{\textsc{Ga}}$(1, 0$\cdot$1) for the exponential power prior;
$\alpha=1,1/(1+\tau)\sim \U(0,1)$ for the generalized
double Pareto prior as suggested by \citet{Armagan11}; and $\tau
\sim \small{\textsc{C}}^+(0,1)$ for the logarithmic prior as was done
for the horseshoe prior in \citet{CarvalhoPolsonScott2010}.

Twenty datasets were generated for each case.  The Bayesian
procedures used 15000 iterations with the first 5000 as burn-ins. In
all cases, the convergence was rapid and the mixing was good; the
autocorrelation of each elements in $\Omega$ died out typically
after 10 lags.  As for the computational cost, the scale mixture of
uniforms and the method of \citet{WongCarter03} were significantly
faster than the \textit{G}-Wishart method. For example,  for model
4, the \textit{G}-Wishart took about 11 hours  for one dataset under
Matlab implementation on a six core 3$\cdot$3 Ghz computer running CentOS
5$\cdot$0 unix ; while the scale mixture of uniforms and the method of
\citet{WongCarter03} took only about 20 and 6 minutes respectively.
The graphical lasso method is just used as a benchmark for calibrating
the Bayesian procedures.
For each dataset, all Bayesian methods were compared to the
graphical lasso method by computing the relative loss; for example,
for the $L_1$ loss, we computed the relative loss as
$L_1(\hat{\Sigma},\Sigma)-L_1(\hat{\Sigma}_\textsc{glasso},\Sigma),$
where $\hat{\Sigma}$ is any Bayes estimator of $\Sigma$ and
$\Sigma_\textsc{glasso}$ is the graphical lasso estimator. Thus, a
negative value indicates that the method performs better relative to
the graphical lasso procedure and a smaller relative loss indicates
a better relative performance of the method.

Table \ref{tab:simu} reports the simulation results. The two approaches based on point mass
priors outperform the continuous shrinkage methods in sparser models
such as model 1, however, they are outperformed in less sparse
configurations such as model 2 and 4. One possible explanation is
that the point mass priors tend to favor sparse models because it
encourages sparsity through a positive prior mass at zero. Finally,
the exponential power with $q=$0$\cdot$2, the generalized double Pareto
and the logarithmic priors have very similar performances --
ranking among top models in all cases. In summary, the experiment
illustrates that these three heavy-tailed priors in the scale
mixture of uniform family are generally indeed good for high dimensional
covariance matrix estimation.


\begin{table}[tbp]
\caption{Summary of the relative $L_1$ and $L_2$ losses for
different models and different methods. Medians are reported while standard errors are in parentheses.
} 
\scriptsize
\begin{tabular}{rlrrrrrrrr}
 &  & \multicolumn{2}{c}{Model 1} & \multicolumn{2}{c}{Model 2} & \multicolumn{2}{c}{Model 3} & \multicolumn{2}{c}{Model 4}\\
 &  & $L_1$   & $L_2$ &$L_1$   & $L_2$&$L_1$   & $L_2$ &$L_1$   & $L_2$  \\
\multirow{5}{*}{n=30}
& W$_{G}$         & -4$\cdot$4 (1$\cdot$3) & -5$\cdot$9 (1$\cdot$4)& -0$\cdot$3 (0$\cdot$7)& -12$\cdot$7 (4$\cdot$6)& -0$\cdot$9 (0$\cdot$7)& 1$\cdot$4 (2$\cdot$5) & -2$\cdot$3 (1$\cdot$9)&  -0$\cdot$0 (0$\cdot$9)\\
& WCK            & -4$\cdot$4 (1$\cdot$0) & -5$\cdot$1 (2$\cdot$3)& -0$\cdot$7 (0$\cdot$6)& -11$\cdot$3 (3$\cdot$8)& -1$\cdot$2 (0$\cdot$6)& 1$\cdot$6 (1$\cdot$5) & -2$\cdot$2 (1$\cdot$0)& 0$\cdot$3 (0$\cdot$5)  \\
 & EP$_{q=1}$     & -2$\cdot$1 (1$\cdot$1)& 2$\cdot$1 (1$\cdot$0)  & -1$\cdot$0 (0$\cdot$8)& -14$\cdot$0 (4$\cdot$7)& -1$\cdot$6 (0$\cdot$7)& -1$\cdot$0 (2$\cdot$2)& -4$\cdot$2 (1$\cdot$2)& -1$\cdot$1 (0$\cdot$5)\\
 & EP$_{q=0\cdot2}$   & -3$\cdot$8 (1$\cdot$1)& -2$\cdot$9 (2$\cdot$1) & -0$\cdot$9 (0$\cdot$8)& -13$\cdot$7 (4$\cdot$9)& -1$\cdot$4 (0$\cdot$7)& -0$\cdot$5 (2$\cdot$5)& -3$\cdot$1 (1$\cdot$1)& -0$\cdot$5 (1$\cdot$3)\\
 & GDP            & -3$\cdot$8 (1$\cdot$1)& -3$\cdot$2 (2$\cdot$2) & -1$\cdot$3 (0$\cdot$7)& -13$\cdot$2 (4$\cdot$3)& -1$\cdot$4 (0$\cdot$7)& -0$\cdot$4 (2$\cdot$3)& -2$\cdot$5 (1$\cdot$7)&  -0$\cdot$4 (0$\cdot$9)\\
 & Log             & -3$\cdot$7 (1$\cdot$1) & -2$\cdot$3 (1$\cdot$4)& -0$\cdot$6 (0$\cdot$8)& -13$\cdot$3 (4$\cdot$9)& -1$\cdot$3 (0$\cdot$6)& -0$\cdot$2 (2$\cdot$5)& -3$\cdot$2 (1$\cdot$1)& -0$\cdot$8 (0$\cdot$9)\\
\\
\multirow{5}{*}{n=100}
 & W$_{G}$        & -1$\cdot$7 (0$\cdot$2) & -3$\cdot$9 (0$\cdot$7)& -0$\cdot$3 (0$\cdot$4)& -0$\cdot$4 (1$\cdot$5) & -0$\cdot$8 (0$\cdot$3)& -1$\cdot$5 (1$\cdot$5)& 0$\cdot$4 (0$\cdot$3)& 0$\cdot$7 (0$\cdot$3)\\
 & WCK            & -1$\cdot$3 (0$\cdot$2) & -2$\cdot$7 (0$\cdot$6)&  -0$\cdot$7 (0$\cdot$3)& -0$\cdot$8 (1$\cdot$1)& -0$\cdot$5 (0$\cdot$2)& 0$\cdot$3 (1$\cdot$4) & 0$\cdot$2 (0$\cdot$3)& 0$\cdot$5 (0$\cdot$3)  \\
 & EP$_{q=1}$     & -0$\cdot$2 (0$\cdot$2)& 0$\cdot$6 (0$\cdot$3)  & -0$\cdot$6 (0$\cdot$3)&  0$\cdot$0 (0$\cdot$8) & -0$\cdot$2 (0$\cdot$3)& 0$\cdot$5 (0$\cdot$5) & -1$\cdot$1 (0$\cdot$3)& -0$\cdot$2 (0$\cdot$1)\\
 & EP$_{q=0\cdot2}$   & -1$\cdot$3 (0$\cdot$2)& -1$\cdot$8 (0$\cdot$3) & -0$\cdot$8 (0$\cdot$3)& -1$\cdot$4 (1$\cdot$2) & -0$\cdot$6 (0$\cdot$2)& -0$\cdot$8 (0$\cdot$7)& -0$\cdot$3 (0$\cdot$3)& 0$\cdot$2 (0$\cdot$2)\\
 & GDP            & -1$\cdot$4 (0$\cdot$2)& -2$\cdot$1 (0$\cdot$4) & -0$\cdot$8 (0$\cdot$4)& -1$\cdot$3 (1$\cdot$1) & -0$\cdot$6 (0$\cdot$2)& -0$\cdot$6 (0$\cdot$6)& -1$\cdot$0 (0$\cdot$3)& -0$\cdot$1 (0$\cdot$1)\\
 & Log             & -1$\cdot$4 (0$\cdot$2) & -1$\cdot$9 (0$\cdot$4)& -0$\cdot$8 (0$\cdot$3)& -1$\cdot$3 (1$\cdot$1) & -0$\cdot$6 (0$\cdot$2)& -0$\cdot$6 (0$\cdot$6)& -0$\cdot$6 (0$\cdot$3)& 0$\cdot$0 (0$\cdot$2)\\
\end{tabular}
\label{tab:simu}
W$_{G}$, \textit{G}-Wishart; WCK, \citet{WongCarter03}; GDP,
generalized double Pareto; EP, exponential power;
Log: logarithmic.
\end{table}

\section{Application to multivariate conditional autoregressive
models}\label{sec:MCAR}
\subsection{Multivariate conditional autoregressive
models based on scale mixture of uniform priors} Multivariate
conditional autoregressive models \citep{gelfand04} constitute a
diverse set of powerful tools for
modeling multivariate spatial random variables at areal unit level.
Let $W=(w_{ij})_{p_r\times p_r}$ be the symmetric proximity matrix
of $p_r$ areal units, $w_{ij} \in \{0,1\}$, and $w_{ii}$ are customarily set to $0$. Then $W$
defines an undirect graph $G_r=(V_r,E_r)$ where an edge $(i,j)\in
E_r$ if and only if $w_{ij}=1$.  Let $w_{i+}=\sum_j w_{ij}$,
$E_W=\textrm{diag}(w_{1+},\ldots,w_{p_r+})$ and  $M =
(m_{ij})=E_W-\rho W$.
Let $X=(x_1,\ldots,x_{p_r})^\T$ denote a $p_r\times p_c$ random matrix
where each $x_i$ is a $p_c$-dimensional vector
corresponding to region $i$. Following \citet{GelfandVounatsou2003},
one popular version of the multivariate conditional autoregressive
models sets the joint distribution of $X$ as
\begin{equation}\label{eq:MCAR:GV}
\textrm{vec}(X) \sim N\{0,(\Omega_c \otimes \Omega_r)^{-1}\},\quad
\Omega_r \mid \rho = E_W-\rho W,\quad \Omega_c \sim
\small{\textsc{W}}(b_c,D_c),
\end{equation}
where $\Omega_r$ is the $p_r\times p_r$ column covariance matrix,
$\Omega_c$ is the $p_c\times p_c$ row covariance matrix, $\rho$
is the coefficient measuring spatial association and is constrained
to be between the reciprocals of the minimum and maximum eigenvalues
of $W$ to ensure that $\Omega_r$ is nonsingular, and $b_c$ and $D_c$ respectively denote
 the degree of freedom and location parameters of a Wishart prior distribution
 for $\Omega_c$. The
joint distribution in (\ref{eq:MCAR:GV}) implies the following
conditional distribution:
$$x_i\mid x_{-i},\rho,\Omega_c \sim \textsc{N}(\sum_{j\in \mathrm{ne}(i) }
\rho \,w_{i+}^{-1} \,x_j,w_{i+}^{-1}\Omega_c),$$ where $\mathrm{ne}(i)$ denotes
the neighbor of region $i$, that is, the set of points satisfying $w_{ij} = 1$. Evidently, the two covariance structures
$(\Omega_r,\Omega_c)$ are crucial in determining the effects of
spatial smoothing. For the matrix $\Omega_c$, direct application of shrinkage
priors can reduce estimation uncertainties as compared to the
conjugate Wishart prior in (\ref{eq:MCAR:GV}). For $\Omega_r$, one
common value of $\rho$ for all $x_i$ may limit the flexibility of the model
 because it assumes the same spatial association for all regions. The recent
work of \citet{DLR2011} uses the \textit{G}-Wishart framework to
provide alternative models. Specifically, the authors recommend the
following extensions for modeling $(\Omega_r,\Omega_c)$:
\begin{equation}\label{eq:MCAR:DLR}
\Omega_r \mid M \sim \W_{G_r}(b_r,M), \quad M \mid \rho = E_W-\rho
W, \quad \Omega_c \sim \W_{G_c}(b_c,D_{c}),
\end{equation}
where the row graph $G_r$ is fixed and obtained from the proximity
matrix $W$, and the column graph $G_c$ is unknown. For both models
in (\ref{eq:MCAR:GV}) and (\ref{eq:MCAR:DLR}), a prior for $\rho$
was chosen to give higher probability mass to values close to 1 to
encourage sufficient spatial dependence. In particular,
\citet{DLR2011} put equal mass on the following 31 values:
$\{$0, 0$\cdot$05, 0$\cdot$1, \ldots, 0$\cdot$8, 0$\cdot$82, \ldots, 0$\cdot$90, 0$\cdot$91, $\ldots$, 0$\cdot$99$\},$. Notice that $\Omega_r$ and $\Omega_c$ are not uniquely identified since, for any $c>0$, $\Omega_c\otimes \Omega_r=(c\,\Omega_c)\otimes (\Omega_r/c)$ \citep{WangWest2009}. We address this by fixing $\Omega_{r,11}=1$.

Using the theory and methods for covariance matrix developed in
Section \ref{sec:smuPre}, we now extend the multivariate conditional
autoregressive models (\ref{eq:MCAR:GV}) using the scale mixture of
uniform distributions. We consider the following two extensions for modeling $\Omega_r\in M^+(G_r)$ and $\Omega_c\in M^+$
\begin{equation} \label{eq:MCAR:WP1}
\Omega_r \mid \rho = E_W-\rho W, \quad p(\Omega_c\mid
\tau) \propto \prod_{i\geq j}g_c(\omega_{c,ij}/\tau_c),
\end{equation}
and
\begin{eqnarray} \label{eq:MCAR:WP2}
 p(\Omega_r) &\propto& \prod_{\{(i,j)\in E_r\} \cup\{i=j\in V_r\}}
g_r(|\omega_{r,ij}-m_{ij}|/\tau_r)\,1_{\{\omega_{r,ij}<0\}}, \quad p(\Omega_c\mid
\tau_c) \propto \prod_{i\geq j}g_c(\omega_{c,ij}/\tau_c).
\end{eqnarray}
The first extension (\ref{eq:MCAR:WP1}) places shrinkage
priors on $\Omega_c$ while leaving the model for $\Omega_r$
unchanged. The second extension (\ref{eq:MCAR:WP2}) further shrinks
$\Omega_r$ towards the matrix $M = E_W-\rho W$ while allowing adaptive spatial
smoothing by not constraining $\Omega_c$ to be controlled by a common
 parameter $\rho$.

One practical advantage of the our model (\ref{eq:MCAR:WP2}) over
the model (\ref{eq:MCAR:DLR}) of \citet{DLR2011} is its flexibility in
incorporating prior knowledge. For example, the similarity of
spatial neighbors implies that the off-diagonal elements of
$\Omega_r$ should be constrained to be negative \citep{gelfand04}.
This point is not addressed by \citet{DLR2011} as their method is
only applicable when the free elements of $\Omega_r$ are not
truncated. In the scale mixture of uniform framework, this important
constraint is easily achieved by truncating each free off-diagonal
element in $\Omega_r$ to be negative when sampling $\Omega_r$. The
functional form of $g_r(\cdot)$ and the shrinkage parameter $\tau_r$
can be pre-specified through careful prior elicitation as follows.
Using the Gibbs sampler in Section \ref{sec:fixtau}, we are able to
simulate from the prior distribution of $\Omega_r$ for fixed
$g_r(\cdot)$ and $\tau_r$. These prior draws allow us to choose
$g_r(\cdot)$ and $\tau_r$ to represent plausible ranges of spatial
associations. To specify these ranges, one guideline can be based on
the model (\ref{eq:MCAR:GV}) for which \citet{GelfandVounatsou2003}
recommended a prior for $\rho$ that favors the upper range of
$\rho\in (0,1)$. In light of this recommendation, we prefer
those $g_r$ and $\tau_r$ that increasingly favor values of
$\omega_{c,ij}$ close to 1 for any $(i,j)\in E_r$ and
$\omega_{c,ii}$ close to $w_{i+}$ for $i\in V_r$. Such choices of
priors integrate prior information about spatial associations and
allow for varying spatial smoothing parameters across different
regions.

\subsection{US cancer data}
Using our model, we analyze the same real data example studied by \citet{DLR2011}  concerning the application of multivariate spatial models
for studying the US cancer mortality rates. The data we analyzed
consists of mortality counts for 10 types of tumors recorded for the
48 mainland states plus the District of Columbia for the year 2000.
The data were collected by the National Center for Health
Statistics. Morality counts below 25 were treated as missing because
they are regarded as unreliable records in cancer surveillance
community. Let $Y_{ij}$ be the number of deaths in state
$i=1,\ldots,p_r=49$ for tumor type $j=1,\ldots,p_c=10.$ Following
\citet{DLR2011}, we considered Poisson multivariate loglinear models
with spatial random effects:
$$Y_{ij} \mid \eta_{ij} \sim \textrm{Poi}(\eta_{ij}),\quad \log(\eta_{ij})=\log(q_i)+\mu_j+X_{ij},$$
where $q_i$ is the population of state $i$, $\mu_j$ is the intercept
of tumor type $j$ and $X_{ij}$ is the zero-mean spatial random
effect associated with state $i$ and tumor $j$ and has the joint
distribution $\textrm{vec}(X) \sim \mathrm{N}\{0,(\Omega_c \otimes
\Omega_r)^{-1}\}$.

We compared the out-of-sample predictive performance of model
(\ref{eq:MCAR:WP1}) and (\ref{eq:MCAR:WP2}) against the model
(\ref{eq:MCAR:GV}) of \citet{GelfandVounatsou2003} and model
(\ref{eq:MCAR:DLR}) of \citet{DLR2011}.  For (\ref{eq:MCAR:GV}) and
(\ref{eq:MCAR:DLR}), we used the same hyper-parameter settings as in
\citet{DLR2011}. For (\ref{eq:MCAR:WP1}), we set $g_c(\cdot)$ to be the
logarithmic density in \eqref{eq:Log} and placed standard half-cauchy prior on
$\tau_c$ in order to expect robust performance for shrinkage
estimation of $\Omega_c$ as was suggested by the simulation study in
Section \ref{sec:cov:simu}. For (\ref{eq:MCAR:WP2}), we let
$g_r(\omega_{r,ij}) \propto
\exp\{-|\omega_{r,ij}-m_{ij}|/\tau_r\}1_{\{\omega_{r,ij}<0\}}$ for
$i=j$ or $(i,j)\in E_r$  so that $\Omega_r$ is centered around
$M=W-E_W$ and the similarity of spatial neighbors is ensured. We did
not choose heavy-tailed distributions for $g_r(\cdot)$ because the
sample size $p_c=10$ is relatively small  for the dimension $p_r=49$
and a heavy-tailed prior can lead to a posterior distribution of
$\omega_{r,ij}$ to be unrealistically small and $\omega_{r,ii}$ to
be unrealistically large. We considered $\tau_r \in \{$0$\cdot$1, 1, 10$\}$
to assess the prior sensitivity. Finally, we modeled $g_c(\cdot)$ as
in model (\ref{eq:MCAR:WP1}).

In order to assess the out-of-sample predictive performance, we
replicated the 10-fold cross-validation experiment of
\citet{DLR2011}. Specifically, we divided the nonmissing counts of
$Y$ into 10 bins. For each bin $i$, we used the samples from the
other 9 bins as observed data and imputed the samples from bin $i$
as missing. To compare different models, we then computed the
predictive mean squared error and mean variance as follows $$\mathrm{MSE}=
{1\over |\{(i,j):Y_{ij}\geq 25\}|}\sum_{\{(i,j):Y_{ij}\geq
25\}}(E(Y_{ij})-Y_{ij})^2,$$ and
$$\mathrm{VAR} = {1\over |\{(i,j):Y_{ij}\geq
25\}|}\sum_{\{(i,j):Y_{ij}\geq 25\}} Var(Y_{ij}),$$  where $E(Y_{ij})$ and $Var(Y_{ij})$ are estimated using the posterior sample mean and variance
based on the output of the analysis of one of the 10 cross-validation datasets in which $Y_{ij}$ are treated as missing. All results were
obtained using a Monte Carlo sample of size 80000 after an initial,
discarded burn-in of 80000 iterations.

\begin{figure}[htbp]
\begin{center}
\begin{tabular}{cc}
  (a) Raw mortality rate & (b) Predicted mortality rate  \\
  \includegraphics[width=3.3in]{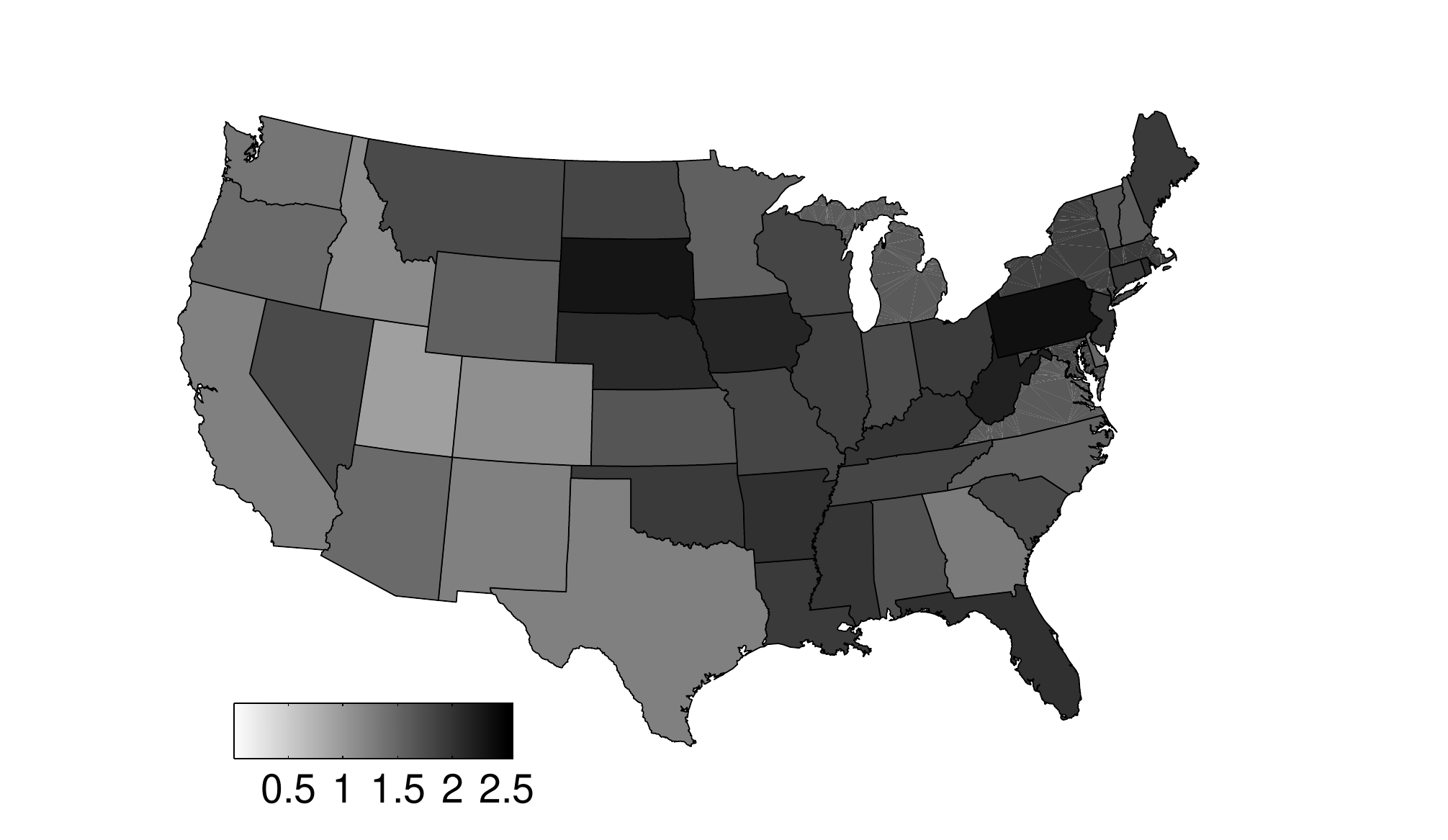}&
  \includegraphics[width=3.3in]{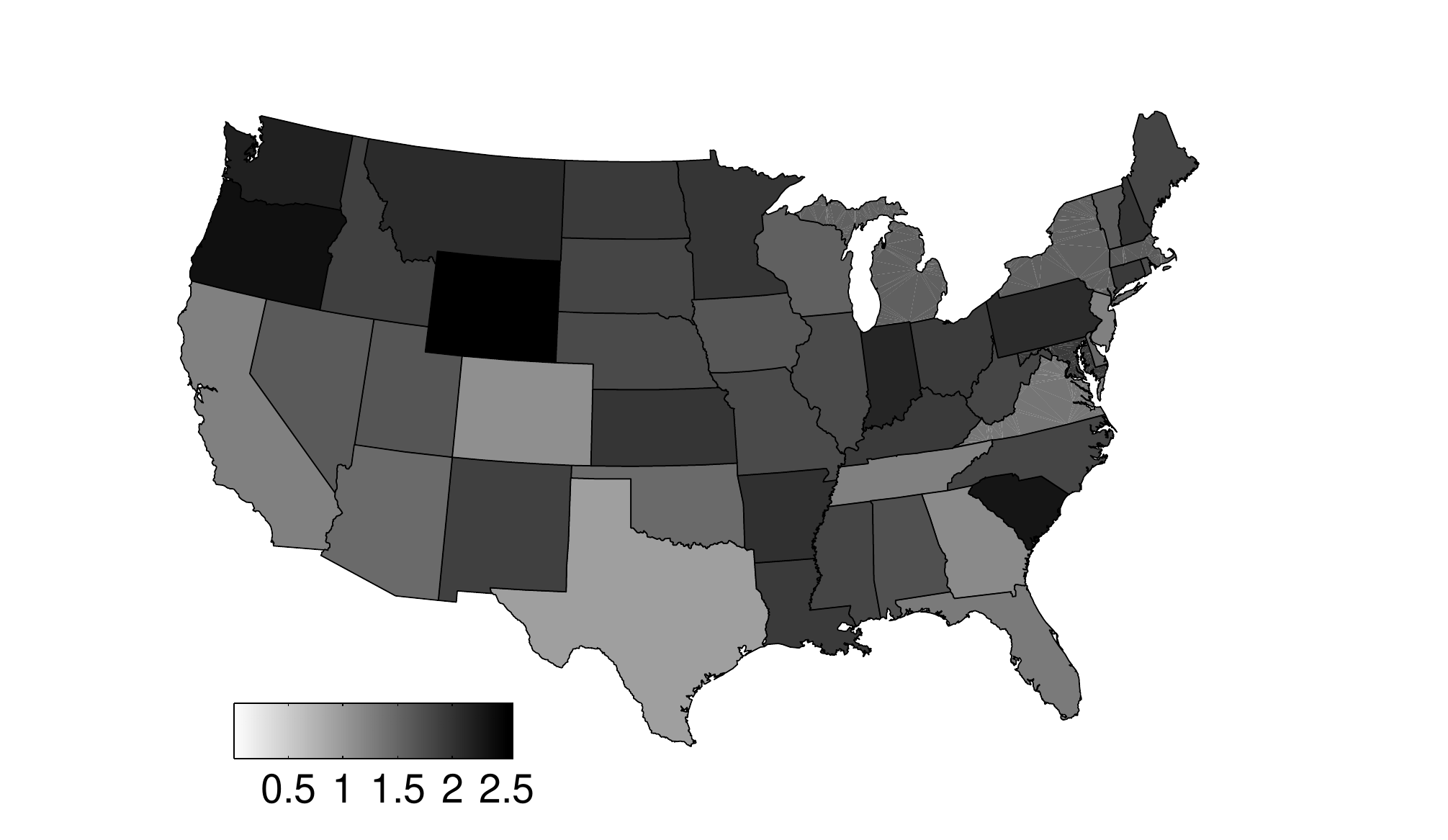}
 \end{tabular}
 \end{center}
 \caption{US cancer mortality map of colon cancer (per 10000 habitants). (a) The raw mortality rate, (b) The predicted mortality rate under model TDE+Log with $\tau_r=1$.\medskip} \label{fig:cancer:usmap}
\end{figure}

Figure \ref{fig:cancer:usmap} shows the raw and predicted morality rate of colon cancer.  Table \ref{tab:mcar} reports the predictive performance as measured
by the mean squared error and mean variance. All methods with
shrinkage priors on $\Omega_c$ improve the prediction over the
standard method using the Wishart prior. Among the shrinkage
methods, the logarithmic prior outperforms the \textit{G}-Wishart
prior.  Allowing $\Omega_r$ to be adaptive by setting
$\tau_r=1$ and 10 can further reduce the mean squared error while
maintaining the same predictive variance  with the common $\rho$
model. Overall, our results suggest that the
models (\ref{eq:MCAR:WP1}) and (\ref{eq:MCAR:WP2}) provide more accurate prediction and
narrower credible intervals than the competing methods for this
dataset.

\begin{figure}[htbp]
\begin{center}
\begin{tabular}{ccc}
  (a) $\tau_r=$0$\cdot$1 & (b) $\tau_r=$1 & (c) $\tau_r=$10\\
  \includegraphics[width=2in]{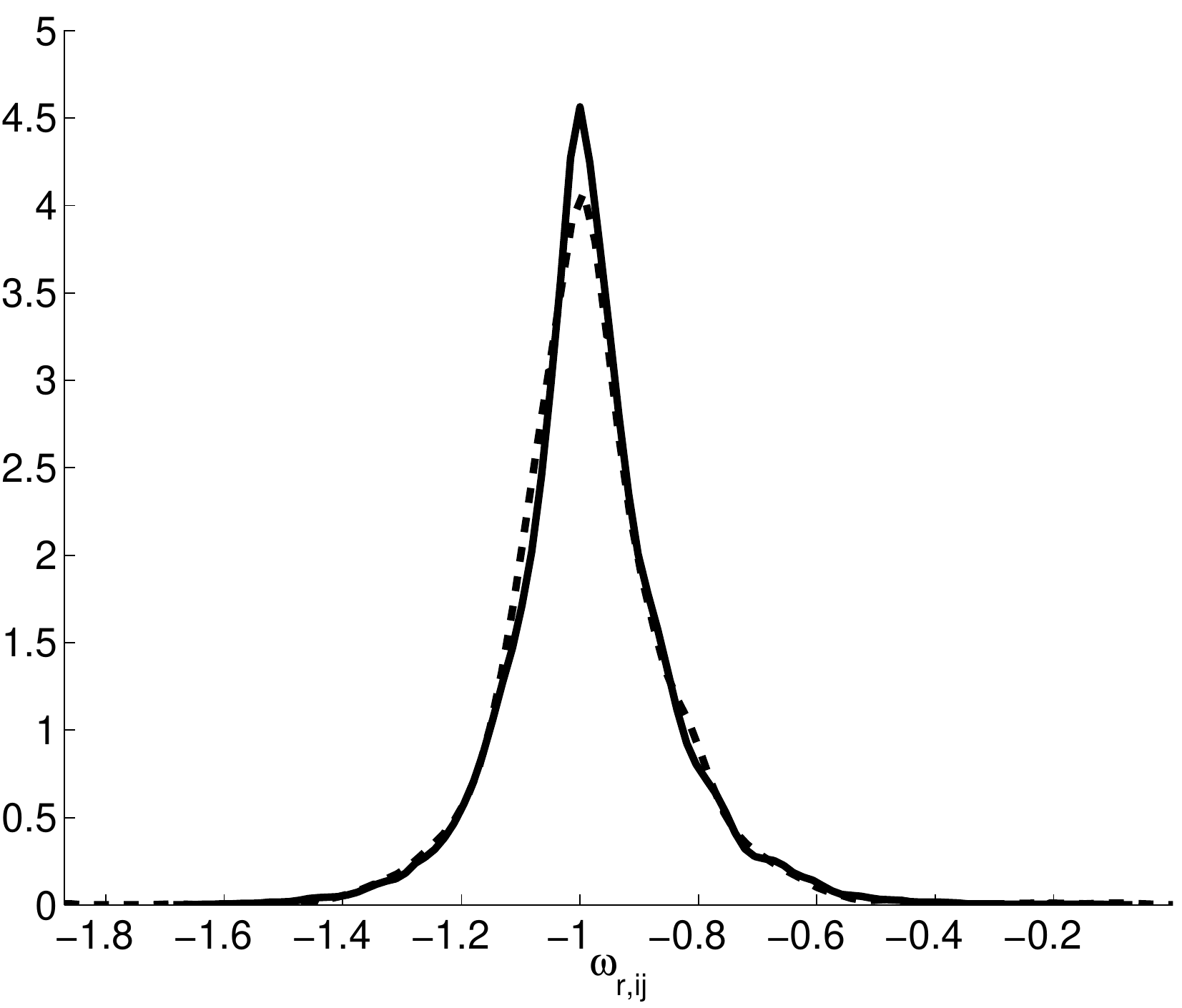}&
  \includegraphics[width=2in]{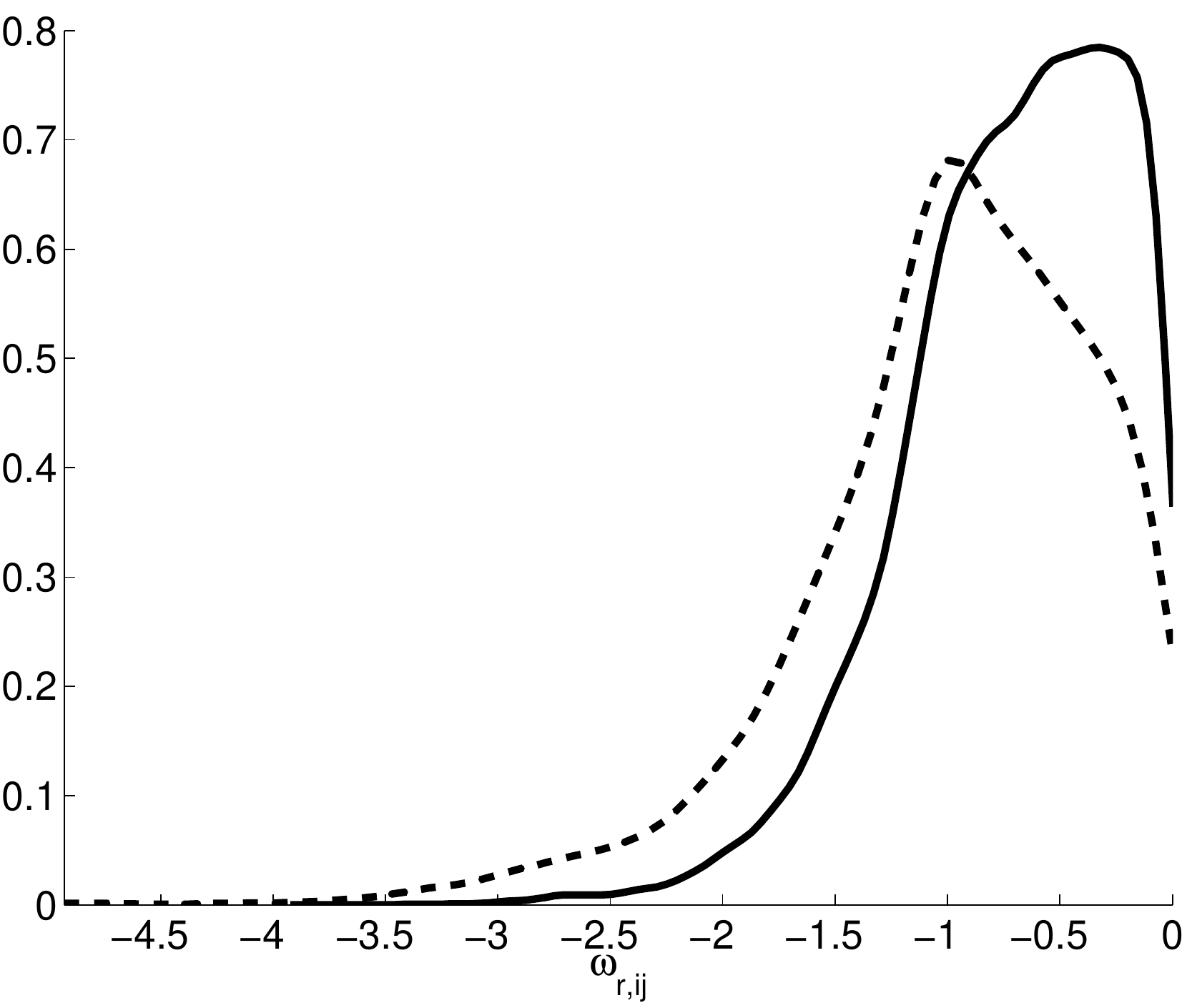}&
  \includegraphics[width=2in]{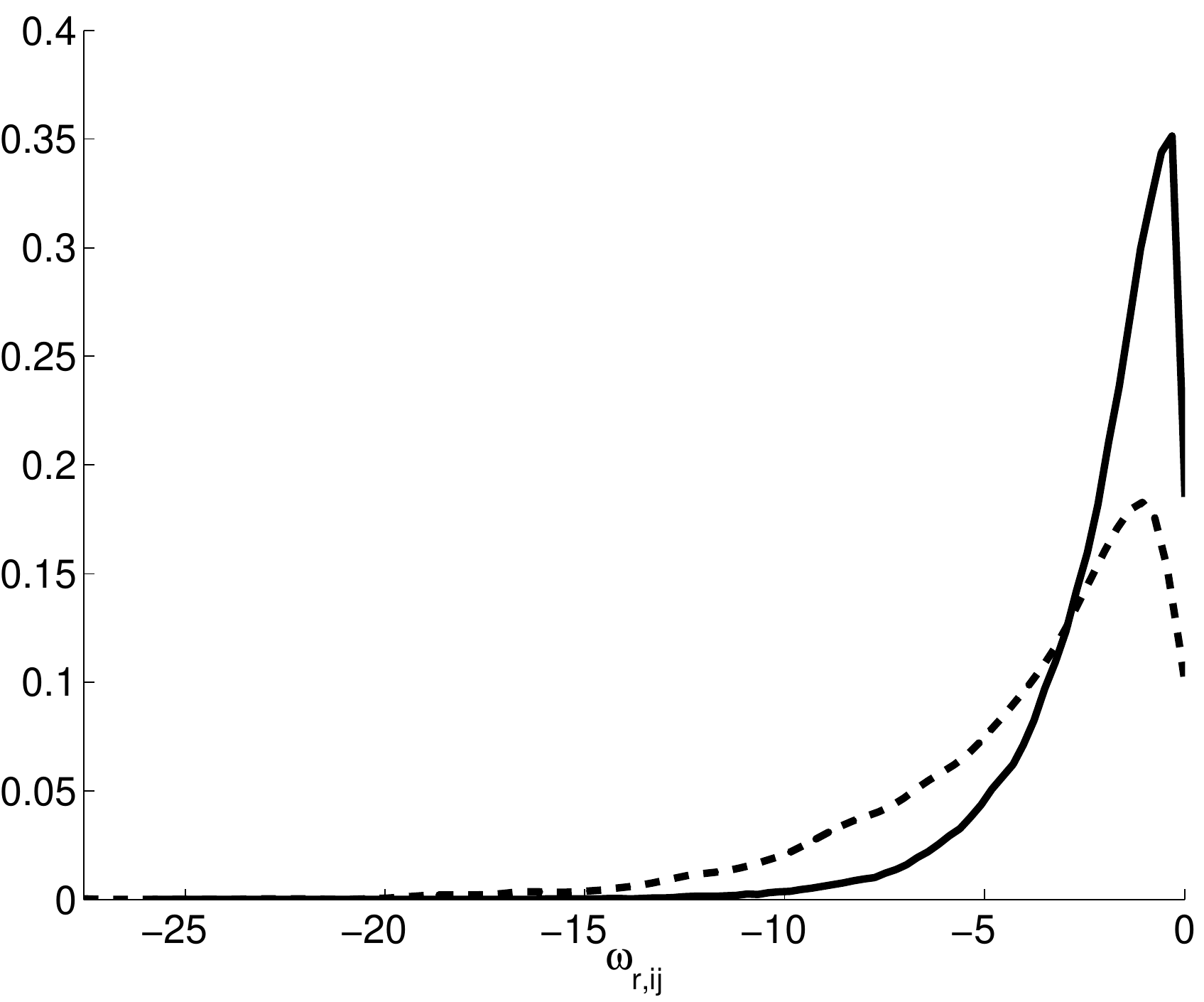}
 \end{tabular}
 \end{center}
 \caption{Marginal prior (dashed lines) and posterior (solid lines) densities of one free off-diagonal element in $\Omega_r$ from
the analysis under model (\ref{eq:MCAR:WP2}) with three different
values of $\tau_r$: (a) $\tau_r=$0$\cdot$1, (b) $\tau_r=$1, (c)
$\tau_r=$10.\medskip} \label{fig:cancer:priorpost}
\end{figure}

To further study the prior sensitivity to the choice of $\tau_r$,
we plotted the marginal prior and posterior densities for the free
off-diagonal element in $\Omega_r$ using samples from the analysis
of the first cross-validation dataset. Figure
\ref{fig:cancer:priorpost} displays the inference for one element
under $\tau_r \in \{$0$\cdot$1, 1, 10$\}$. Clearly, the marginal posterior
distribution depends on the choice of $\tau_r$. This is not surprising
because the sample size is small compared to the dimension of
$\Omega_r$. The case $\tau_r=1$ and 10 seems to perform well in
this example because the marginal posterior distribution is
influenced by the data.  The case $\tau_r=$0$\cdot$1 appears to be too tight
and thus is not largely influenced by the data.

On the computing time, the Matlab implementation of model
(\ref{eq:MCAR:WP2}) took about 4 hours to complete the analysis of
one of the ten cross-validation datasets, while model
(\ref{eq:MCAR:DLR}) of \citet{DLR2011}  took about 4 days.
Additionally, \citet{DLR2011} reported a runtime of about 22 hours
on a dual-core 2$\cdot$8 Ghz computer under C++ implementation for a
similar dataset of size $p_r=49$ and $p_c=11$. As mentioned above,
our models based
on the scale mixture of uniforms are not only more flexible but also
more computationally efficient.

\begin{table}[tbp]
\caption{Predictive mean squared error and variance in 10-fold cross-validation predictive performance in the cancer mortality example.}
\footnotesize
\begin{tabular}{ccccccc}
            &GV    & DLR   & Common $\rho$+Log  & \multicolumn{3}{c}{TDE+Log} \\
            &      &       &      & $\tau_r$=10 & $\tau_r$=1 & $\tau_r$=0$\cdot$1\\
       MSE  & 3126 &  2728 & 2340 & 2238             & 2187        & 2359 \\
       VAR  & 9177 &  6493 & 3814 & 3850             & 3810        & 3694
\end{tabular}
 \label{tab:mcar}
\\
\\
 GV: the non-shrinkage model
(\ref{eq:MCAR:GV}) of \citet{GelfandVounatsou2003}; DLR: model
(\ref{eq:MCAR:DLR}) of \citet{DLR2011}; Common $\rho$+Log: model
(\ref{eq:MCAR:WP1}) under common $\rho$ for $\Omega_r$ and
logarithmic prior for $\Omega_c$; TDE+Log: model
(\ref{eq:MCAR:WP2}) under truncated double-exponential prior for
$\Omega_r$ with fixed but different $\tau_r$ and logarithmic
prior for $\Omega_c$.
\end{table}

\section{Shrinkage prior for linear regression
models}\label{sec:reg}
In this section we briefly investigate the properties of the shrinkage prior constructed from scale mixture
of uniforms for the linear regression models. Recently, shrinkage
estimation for  linear models have received a lot of attention
\citep{ParkCasellla2008,GriffinBrown10,Armagan11}
all of which proceed via the
scale mixture of normals.  \citet{Walker97} and \citet{Walker98} were among the first to use the scale mixture of uniform priors  for regression models. However, they used this family only for modeling the measurement errors and deriving the corresponding Gibbs sampler.  To the best of our knowledge, we are the first to investigate the scale mixture of uniforms as a class of shrinkage priors for regression coefficients. When this paper was nearing completion we were notified of a similar approach in the very recent work \citet{PolsonScott2011} in which the authors independently propose a similar construction based on mixtures of Bartlett-Fejer kernels for the bridge regression and proceed via a result similar
to Theorem \ref{therm1}.

Consider the following version of a regularized Bayesian linear
model where the goal is to sample from the posterior distribution
\begin{equation*}
p(\beta \mid \sigma,\tau,Y) \propto \exp\{-{1\over
2\sigma^2}(Y-X\beta)^\T(Y-X\beta)\} \prod^p_{j=1} g({\beta_j \over
\sigma \tau})
\end{equation*}
where $g(\cdot)$ is the shrinkage prior and $\tau$ is the global
shrinkage parameter.  Theorem \ref{therm1} suggests we can introduce
latent variable $t=\{t_1,\ldots,t_p\}$ such that the joint posterior
of $(\beta,t)$ is given by:
\begin{equation*}
p(\beta, t \mid \sigma,\tau,Y) \propto \exp\{-{1\over
2\sigma^2}(Y-X\beta)^\T(Y-X\beta)\} \prod^p_{j=1}
\{-g^\prime(t_j)\,1_{\{\sigma\tau t>|\beta_j|\}}\}
%
\end{equation*}
The Gibbs samplers are then implemented by (a) simulating $\beta_j$
from a truncated normal for each $j$, and (b) block simulating
$\{t_1,\ldots,t_p\}$ from using the conditional cumulative distribution
function in Theorem \ref{therm:sampler}.

We compare the posterior mean estimators under the exponential power
prior with $q=$0$\cdot$2 and the logarithmic prior to the posterior
means corresponding to several other existing priors. These two shrinkage
 priors are
interesting because the exponential power prior is the Bayesian
analog of the bridge regression \citep{ParkCasellla2008} and is challenging for fully
posterior analysis using the scale mixture of normals and relatively
unexplored before, and the logarithmic prior is a new prior that
resembles the class of horseshoe priors that are shown to have
some advantages over many existing approaches \citep{CarvalhoPolsonScott2010}.

We use the setting of simulation experiments considered in
\citet{Armagan11}. Specifically,  we generate $n=50$ observations from $y=x^\T\beta+\epsilon,\epsilon \sim \N(0,3^2)$,
where $\beta$ has one of the following five configurations: (i) $\beta=$(1,1,1,1,1,0,0,0,0,0,0,0,0,0,0,0,0,0,0,0)$^\T$,
(ii) $\beta$(3,3,3,3,3,0,0,0,0,0,0,0,0,0,0,0,0,0,0,0)$^\T$, (iii) $\beta=$(1,1,1,1,1,0,0,0,0,0,1,1,1,1,1,0,0,0,0,0)$^\T$, (iv) $\beta=$(3,3,3,3,3,0,0,0,0,0,3,3,3,3,3,0,0,0,0,0)$^\T$, (v) $\beta=$(0$\cdot$85, $\ldots$, 0$\cdot$85)$^\T$,
and $x=(x_1,\ldots,x_p)^\T$ has one of the following two scenarios: (a) $x_{j}$ are independently  and identically distributed standard normals, (b) $x$ is a multivariate normal with $E(x)=0$ and
$\mathrm{cov}(x_j,x_{j'})=$ 0$\cdot$5$^{|j-j'|}$. The variance is
assumed to have the Jeffrey's prior $p(\sigma^2)\propto 1/\sigma^2$.
The global shrinkage parameter is assumed to have the conjugate
$\tau^{-q} \sim \rm{Ga}(1,1)$ for the exponential power prior with
$q=$0$\cdot$2, and $\tau \sim \C^+(0,1)$ for the logarithmic prior.
Model error is calculated using the Mahalanobis distance $(\hat{\beta}-\beta)^\T
\Sigma_X(\hat{\beta}-\beta)$ where $\Sigma_X$ is the covariance
matrix used to generate $X$.

Table \ref{tab:reg} reports the median model errors and the
bootstrap standard errors based on 100 datasets for each case.
Results for cases other than the exponential power prior with $q=$0$\cdot$2 and the
logarithmic prior are based on the reported values of
\citet{Armagan11}. Except for model (iii) and (v)  in the correlated
predictor scenario, the exponential power prior with $q=1$ is
outperformed by other methods. The performances of the exponential
power prior with $q=$0$\cdot$2 and the logarithmic prior are comparable
with those of the generalized Pareto and the horseshoe priors.

\begin{table}[tbp]
\caption{Summary of model errors for the simulation study in the
regression analysis of Section \ref{sec:reg}. Median model errors
are reported; bootstrap standard errors are in parentheses.}
\scriptsize
\bigskip
\begin{tabular}{lccccccccccc}
& \multicolumn{5}{c}{(a) $x_j$ independent}& &\multicolumn{5}{c}{(b) $x_j$ correlated}  \\
&(i) &(ii)&(iii)&(iv)&(v) & &(i) &(ii)&(iii)&(iv)&(v)\\
 GDP$^a$ &  2$\cdot$7 (0$\cdot$1)& 2$\cdot$2 (0$\cdot$2)& 4$\cdot$0 (0$\cdot$2) &3$\cdot$8 (0$\cdot$2) & 5$\cdot$7 (0$\cdot$3) &
&  2$\cdot$1 (0$\cdot$1)& 2$\cdot$1 (0$\cdot$1)& 3$\cdot$2 (0$\cdot$1) &4$\cdot$2 (0$\cdot$3) & 4$\cdot$4 (0$\cdot$1) \\
 GDP$^b$ &  2$\cdot$8 (0$\cdot$2)& 2$\cdot$1 (0$\cdot$2)& 4$\cdot$6 (0$\cdot$2) &3$\cdot$8 (0$\cdot$2) & 7$\cdot$0 (0$\cdot$2) &
&  1$\cdot$9 (0$\cdot$1)& 2$\cdot$0 (0$\cdot$1)& 3$\cdot$3 (0$\cdot$2) &4$\cdot$2 (0$\cdot$2) & 4$\cdot$7 (0$\cdot$1) \\
 GDP$^c$ &  2$\cdot$6 (0$\cdot$1)& 2$\cdot$4 (0$\cdot$2)& 4$\cdot$4 (0$\cdot$2) &4$\cdot$0 (0$\cdot$2) & 6$\cdot$5 (0$\cdot$2) &
&  1$\cdot$9 (0$\cdot$1)& 2$\cdot$2 (0$\cdot$1)& 3$\cdot$1 (0$\cdot$2) &4$\cdot$3 (0$\cdot$2) & 4$\cdot$3 (0$\cdot$1) \\
 HS &  2$\cdot$7 (0$\cdot$1)& 2$\cdot$1 (0$\cdot$2)& 4$\cdot$8 (0$\cdot$2) &3$\cdot$8 (0$\cdot$2) & 7$\cdot$3 (0$\cdot$2) &
&  2$\cdot$0 (0$\cdot$1)& 2$\cdot$0 (0$\cdot$1)& 3$\cdot$3 (0$\cdot$2) &4$\cdot$3 (0$\cdot$2) & 4$\cdot$6 (0$\cdot$1) \\
EP$_{q=1}$ &  3$\cdot$2 (0$\cdot$1)& 4$\cdot$0 (0$\cdot$3)& 5$\cdot$1 (0$\cdot$3) &4$\cdot$9 (0$\cdot$3) & 7$\cdot$3 (0$\cdot$5) &
 & 2$\cdot$1 (0$\cdot$1)&  2$\cdot$8 (0$\cdot$2)& 2$\cdot$8 (0$\cdot$1) &4$\cdot$2 (0$\cdot$3) & 3$\cdot$5 (0$\cdot$2)\\
EP$_{q=0\cdot2}$ &  2$\cdot$5 (0$\cdot$1)& 2$\cdot$0 (0$\cdot$1)& 4$\cdot$7 (0$\cdot$1) &3$\cdot$9 (0$\cdot$3) & 7$\cdot$3 (0$\cdot$3)&
 & 2$\cdot$0 (0$\cdot$1)&  2$\cdot$1 (0$\cdot$1)& 3$\cdot$2 (0$\cdot$1) &3$\cdot$9 (0$\cdot$1) & 5$\cdot$4 (0$\cdot$2)\\
Log &  2$\cdot$5 (0$\cdot$1)& 2$\cdot$5 (0$\cdot$2)& 4$\cdot$5 (0$\cdot$2) &4$\cdot$5 (0$\cdot$2) & 6$\cdot$4 (0$\cdot$4) &
 & 2$\cdot$0 (0$\cdot$1)&  2$\cdot$4 (0$\cdot$1)& 3$\cdot$0 (0$\cdot$1) &4$\cdot$3 (0$\cdot$1) & 4$\cdot$6 (0$\cdot$2)\\
\end{tabular} 
\label{tab:reg}
GDP$^{a,b,c}$,
three recommended Generalized double Pareto priors in \citet{Armagan11}; HS, horseshoe; EP, exponential power;
Log, logarithmic.
\end{table}

\section{Conclusion}
The scale mixture of uniform prior provides a unified framework for shrinkage estimation of covariance matrices for a wide class of prior distributions.  Further research on the scale mixture of uniform distributions is of interest in developing theoretical insights as well as computational advances in shrinkage prior estimation for Bayesian analysis of covariance matrices and other related models. One obvious next step is to investigate the covariance selection models that encourage exact zeros on a subset of elements of $\Omega$ under the scale mixture uniform priors.  Such extensions can potentially combine the flexibility of the scale mixture of uniform priors and the interpretation of the graphs implied by exact zero elements. Another interesting research direction is the generalization of the basic random sampling models to dynamic settings that allow the covariance structure to be time-varying. Such models are useful for analyzing high-dimensional time series data encountered in areas such as finance and environmental sciences.  We are current investigating these extensions and we expect  the Gibbs sampler developed in Section \ref{sec:fixtau} to play a key role in model fitting in these settings.
\section*{Acknowledgements}
The authors thank Abel Rodriguez and James G. Scott for very useful conversations and references. NSP gratefully acknowledges the NSF grant DMS 1107070.


\appendix
\section*{Appendix}
\subsection*{Details of sampling algorithm in Section \ref{sec:fixtau}}
The joint distribution of $(l_{12},d_1,d_2)$ is: $$ p(
d_1,d_2,l_{21} \mid -) \propto d_1^{n/2+1}d_2^{n/2} \exp[-{1\over
2}\tr \{s_{11}d_1+s_{22}(l_{21}^2d_1+d_2)+2s_{21}d_1l_{21}\} ]\,1_{\{
\Omega_{e,e}\in \mathcal{T} \}} \label{cov:post2}.
$$
Clearly, the full conditional distribution for $d_1$, $d_2$  and $l_{21}$ are given by
$$d_1 \sim \small{\textsc{Ga}}\{{n/2+2},(s_{11}+s_{22}l_{21}^2+2s_{21}l_{21})/2\} \,1_{\{ \Omega_{e,e}\in \mathcal{T} \}}\,\;,$$ $d_2 \sim \small{\textsc{Ga}}({n/2+1},{s_{22}/2}
)\, 1_{\{\Omega_{e,e}\in \mathcal{T}\}}$ and $l_{21}\sim \small{\textsc{N}}\{s_{21}/s_{22}, 1/(s_{22}d_1) \}\,1_{\{\Omega_{e,e}\in \mathcal{T}\}}$, respectively. To identify the truncated region $\mathcal{T}$, recall $$\Omega_{e,e}=A+B,\quad A = \left(\begin{array}{cc}
                               d_1 & d_1 l_{21}\\
                               d_1 l_{21} & d_1 l_{21}^2 +d_2 \\
                             \end{array}
                           \right), \quad B = \left(\begin{array}{cc}
                               b_{11} & b_{12} \\
                               b_{21} & b_{22} \\
                             \end{array}
                           \right).$$
The set
$\mathcal{T} = \{|\omega_{ij}|<t_{ij}\}\cap\{|\omega_{ii}|<t_{ii}\}
\cap \{|\omega_{jj}|<t_{jj}\}$  can be written as
\begin{equation} \{|d_1 + b_{11}|<t_{ii}\} \cap\{|d_1 l_{21}+b_{21}|<t_{ij}\}
\cap \{|d_1 l_{21}^2 +d_2 +b_{22}|<t_{jj}\} \label{eq:trRegion}.\end{equation}
Given $\{B,t_{ii},t_{ij},t_{jj}\}$, (\ref{eq:trRegion}) gives straightforward expressions for the truncated region of each variable in  $(d_1,d_2,l_{21})$ conditional on the other two.

Sampling a univariate truncated normal can be carried out
efficiently using the method of \citet{Robert1995}, while sampling a
truncated gamma is based on the inverse cumulative distribution function method.

\bibliography{BayesMixUnif}
\bibliographystyle{biometrika}
\end{document}